\documentclass[pra,amsmath,amssymb]{revtex4}
\usepackage{amsfonts}
\usepackage{amssymb}
\usepackage{graphicx}
\usepackage{subfigure}
\usepackage{color}
\usepackage{ulem}

\newcommand{\be}{\begin{equation}}
\newcommand{\ee}{\end{equation}}
\newcommand{\bea}{\begin{eqnarray}}
\newcommand{\eea}{\end{eqnarray}}
\newcommand{\ba}{\begin{array}}
\newcommand{\ea}{\end{array}}

\begin{document}

\title{ Characterization of Unruh Channel in the context of Open Quantum Systems}

\author{Subhashish Banerjee}
\email{subhashish@iitj.ac.in}
\affiliation{Indian Institute of Technology Jodhpur, Jodhpur 342011, India}

\author{Ashutosh Kumar Alok}
\email{akalok@iitj.ac.in}
\affiliation{Indian Institute of Technology Jodhpur, Jodhpur 342011, India}

\author{S. Omkar}
\email{omkar.shrm@gmail.com }
\affiliation{Indian Institute of Science Education and Research, Thiruvananthapuram, India }

\author{R. Srikanth}           \email{srik@poornaprajna.org}
\affiliation{Poornaprajna    Institute    of   Scientific    Research,
  Sadashivnagar, Bengaluru- 560080, India}

\begin{abstract}
We show  through the Choi  matrix approach  that the effect  of Unruh
acceleration on  a qubit is  similar to  the interaction of  the qubit
with a vacuum  bath, despite the finiteness of  the Unruh temperature.
Thus, rather counterintuitvely, from the perspective of decoherence in
this  framework,  the  particle  experiences  a  vacuum  bath  with  a
temperature-modified interaction strength, rather than a thermal bath.
 We investigate how this ``relativistic decoherence'' is modified
  by the presence of  environmentally induced decoherence, by studying
  the degradation of quantum  information, as quantified by parameters
  such as  nonlocality, teleportation  fidelity, entanglement,  coherence
  and   quantum   measurement-induced  disturbance   (a   discord-like
  measure).  Also studied are the performance parameters such as gate and channel fidelity. 
  We  highlight  the  distinction  between  dephasing  and
  dissipative environmental  interactions, by considering  the actions
  of quantum non-demolition and squeezed generalized amplitude damping
  channels, respectively, where, in particular, squeezing is shown to be 
  a useful quantum resource. 
\end{abstract}

\maketitle

\section{Introduction}

Quantum information is emerging in the forefront of processing and harnessing
of information, using the advantages of quantum mechanics, as well as becoming an essential
ingredient towards understanding the fundamental aspects of nature. Relativistic
quantum information is a natural quest in this direction \cite{czachor1997,peres1,caban1,Banerjee:2014vga,Alok:2014gya,Banerjee:2015mha, fuentes2010}. 
In this context, Unruh effect \cite{Dav-Unr,crispino} which predicts
thermal effects from observing uniform acceleration of the  Minkowski vacuum  has attracted intense interest \cite{AM03,alsing,Tian2012,
Lee:2014kaa,Peres:2002wx}.

Experimental progress in this direction is now attracting considerable attention from the community. Circuit quantum electrodynamics (cQED), using  
Superconducting Quantum Interferometric Devices (SQUIDs), is a promising effort in this direction. Here tuneable
boundary conditions are possible, corresponding to mirrors moving at speeds close to the speed of light in the medium. This was used to 
experimentally simulate the scenario of dynamical Casimir effect \cite{nori}, hitherto belonging to the regime of quantum field theory. This
paved the way for investigations into various facets of relativistic quantum information. Thus, for example, there have been investigations into
relativistic quantum teleportation \cite{fuentes:relteleport}, entanglement generation and gate operations using superconducting resonators \cite{anderson, bruschi13},
entanglement generation via the dynamical Casimir effect using SQUID \cite{felicetti14, sousa15}, relativistic effects with superconducting qubits \cite{felicetti15}
and universal quantum computation, using continuous variable Gaussian cluster states, through relativistic motion of cavity mirrors \cite{fuentesscreport}. Further,
there have been investigations of twin paradox in superconducting circuits \cite{lindkvist} as well as using the geometric phase to detect the Unruh temperature
at accelerations small enough to be experimentally feasible \cite{fuentesGP}. There have also been investigations into the quantum metrology aspects of Unruh effect 
\cite{aspachs2010, yao2014, sbfisher2015}.

This sets the scene for the present investigation where the effect of environmentally induced decoherence, 
an inevitable attribute for an experimental realization, on various facets of quantum  information, as quantified by parameters
such as  nonlocality, teleportation  fidelity, entanglement,  coherence and   quantum   measurement-induced  disturbance (a   discord-like  measure) are studied.   
We  highlight  the  distinction  between  dephasing  and dissipative environmental  interactions, by considering  the actions
of quantum non-demolition (QND) and squeezed generalized amplitude damping (SGAD) channels, respectively. The QND channel is a 
purely quantum effect incorporating decoherence without dissipation \cite{sbrg}, tracing its roots to gravitational wave detection \cite{braginsky,caves}.
The SGAD \cite{sbsrik,gate} is a very general dissipative channel of the Lindblad class and incorporates the well known amplitude damping (AD) and generalized
amplitude damping (GAD) channels as limiting cases. In \cite{us}, the pure Unruh channel for the Dirac qubit, that is, the Unruh channel without
any external influence, was characterized information theoretically by constructing its Kraus operators. Here, the Unruh channel is quantified taking
into account possible ambient external influences.  Useful  parameters characterizing channel performance are the gate fidelity \cite{Mag11} as well
as the average gate fidelity \cite{BMD+02,Nie02,gate}. They represent how well  a  (noisy)  gate  performs  the  operation  it  is  supposed  to implement. How well 
a gate preserves the distinguishability of states  is captured  by another channel  performance parameter, the channel  fidelity, introduced  in \cite{sbsrik}. These 
channel parameters are applied here to the Unruh channel, both with and without external influences.  

The plan of the work is as follows. In Section II, we briefly discuss the Unruh channel for the Dirac qubit with emphasis towards its information theoretic 
characterization. This is followed by studying how external influences, characterized by parameters like temperature, squeezing and evolution time, effect various
features of quantum information, such as nonlocality, entanglement, teleportation fidelity and  measurement-induced  disturbance. Next, the average gate and channel
fidelities are studied in order to gain insight into the nature of the Unruh channel. Quantum coherence, characteristic of a quantum operation, 
as a resource theory has attracted a lot of attention in recent times. Mixing, which is inevitable with evolution, will result in a degradation of coherence.
In order to have an operational etimation of the utility of a quantum task, it is imperative to have an understanding of the trade-off between the 
two \cite{coh1,coh3,coh4}. This is done for the Unruh channel, pure as well as in the presence of ambient effects, in the penultimate section. We then make our 
conclusions.

\section{Invitation to Unruh effect}

We consider two  observers, Alice (A) and Rob (R)  sharing a maximally
entangled initial  state of two  qubits at  a point in  flat Minkowski
spacetime. Then Rob moves with  a uniform acceleration and Alice stays
stationary.  Moreover, we assume that  the observers are equipped with
detectors that are sensitive only  to their respective modes and share
the    following    maximally    entangled    initial    state:    \be
|\psi\rangle_{A,R}=\frac{|00\rangle_{A,R}+|11\rangle_{A,R}}{\sqrt{2}}.
\label{eqn:unruh}
\ee

\begin{figure}
\begin{center}
\includegraphics[width=0.5\linewidth]{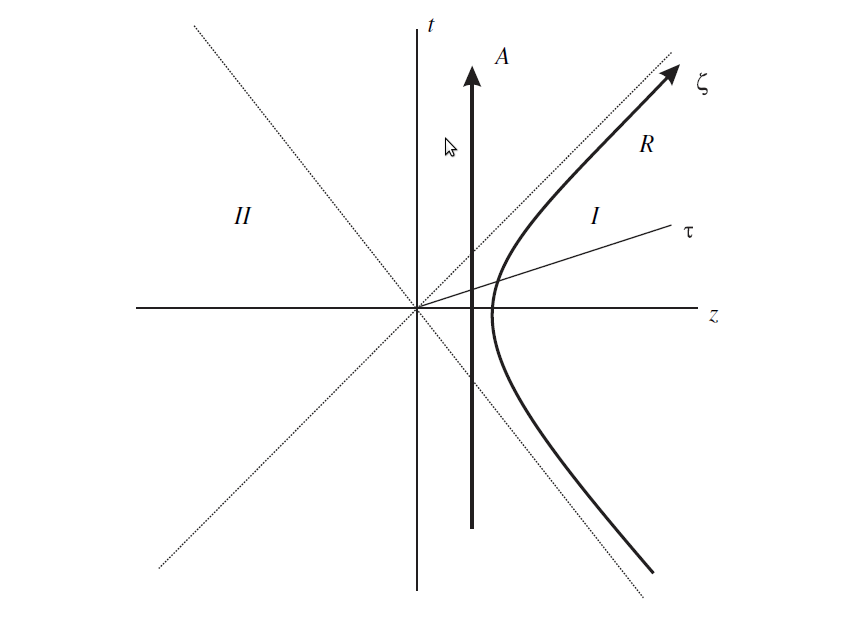}
\caption{Minkowski space}
\label{fig:min}
\end{center}
\end{figure}

Assuming that R gets uniformly accelerated with acceleration $a$, from
R's frame  the Minkowski  vacuum state  is a  two mode  squeezed state
\cite{alsing}:    \be    |0\rangle_M=\cos(r)|0\rangle_I|0\rangle_{II}+
\sin(r)|1\rangle_I|1\rangle_{II},
\label{eq:unruh1}
\ee
and the excited state is 
\be 
|1\rangle_M=\cos(r)|1\rangle_I|1\rangle_{II},
\label{eq:unruh2}
\ee where $\cos(r)=\frac{1}{\sqrt{e^{-\frac{2\pi\omega c}{a}+1}}}$ and
$\omega$ is  Dirac particle acceleration  with $c$ being the  speed of
light in vacuum.

From the Fig.~(\ref{fig:min}),  it can be seen that the  regions I and
II are  causally disconnected.  The state  in Eq.~(\ref{eqn:unruh}) has
contribution  from the  regions I  and  II.  Since  these regions  are
disconnected, the mode corresponding to II can be traced out to obtain
the      following      density       matrix      \be      \rho_{A,I}=
\frac{1}{2}\left[\cos^2(r)|00\rangle\langle
  00|+\cos(r)(|00\rangle\langle11|+|11\rangle\langle
  00|)+\sin^2(r)|01\rangle\langle 01| +|11\rangle\langle 11|\right].
\label{un:matrix}
\ee

Consider the  maximally entangled two  mode state in which  the second
mode  is  Unruh   accelerated.   The  state  is   represented  by  \be
\rho_u=\frac{1}{2}\left(
\begin{array}{cclr}
\cos^2r&0&0&\cos r\\
0&\sin^2r&0&0\\
0&0&0&0\\
\cos r&0&0&1
\end{array}\right),
\ee
where $\sin(r)=\frac{1}{1+e^{\frac{2\pi \omega c}{a}}}$.

Leaving out the factor $1/2$ in $\rho_u$,  the corresponding Choi matrix \cite{choi,debbie} is $|i\rangle\langle i|\mathcal{E}(|j\rangle\langle j|)$,
from which the  Kraus representation of the Unruh channel can be obtained. Following the procedure for constructing Kraus operators we have 
\be
K^u_1=\left(
\begin{array}{cclr}
\cos r&0\\
0&1
\end{array}\right),~~
K^u_2=\left(
\begin{array}{cclr}
0&0\\
\sin r&0
\end{array}\right).
\label{eq:ukraus}
\ee

The above Kraus operators are  similar to 
\be
K_1=\left(
\begin{array}{cclr}
\sqrt{1-\gamma}&0\\
0&1
\end{array}\right),~~
K_2=\left(
\begin{array}{cclr}
0&0\\
\sqrt{\gamma}&0
\end{array}\right),
\ee which represents  the dissipative interaction of a  qubit with the
vacuum bath,  i.e., an  amplitude damping  channel \cite{sbsrik,gate}.
Here   $\gamma=\frac{1}{e^{\frac{\omega}{k_B   T}}}$.   Thus,   rather
surprisingly, from the viewpoint of decoherence, the qubit experiences
a vacuum bath, with the role  of Unruh temperature being to modify the
interaction strength, rather than to serve as bath temperature.

To see what acceleration $a$ simulates the temperature $T$, let us equate
\be
\frac{1}{1+e^{\frac{2\pi \omega c}{a}}}=\frac{1}{e^{\frac{\omega}{k_B T}}}.
\ee

Thus we have the relation between $T$ and $a$ as
\be
\frac{\hbar\omega}{k_B T}=\log\left(1+e^{\frac{2\pi\hbar \omega c}{a}}\right).
\ee
For small $a$, 1 in the above equation can be ignored and we have the relation
\be
a=2\pi k_B c T.
\ee

For $a\longrightarrow\infty$, $1+e^{\frac{2\pi\hbar \omega c}{a}}\longrightarrow 2$ and the corresponding temperature is 
\be
T_{\infty}=\frac{\hbar\omega}{k_B \log2}.
\ee
$T_{\infty}$ is the maximum temperature that can be simulated by Unruh effect. Since the maximum temperature is 
never infinite, the asymptotic state due to Unruh acceleration is never a maximally mixed state and entanglement is seen to 
survive \cite{us}. In contrast, the  qubit interacting with a thermal bath can become maximally mixed for infinite bath temperatures.

Once we have the Kraus operators of the Unruh channel, we can calculate its effect on a qubit in 
pure state given by
\be
\rho=\left(\begin{array}{clclr}
\cos^2(\theta/2)& e^{i\phi}\cos(\theta/2)\sin(\theta/2)\\
e^{-i\phi}\cos(\theta/2)\sin(\theta/2)& \sin^2(\theta/2)
\end{array}\right).
\ee
The action of the Unruh channel on the state $\rho$ is
\be
\mathcal{E}_u(\rho)= \left(\begin{array}{clclr}
\cos^2r\cos^2(\theta/2)& \cos r e^{i\phi}\cos(\theta/2)\sin(\theta/2)\\
\cos r e^{-i\phi}\cos(\theta/2)\sin(\theta/2)& \sin^2r\cos^2(\theta/2) +\sin^2(\theta/2)
\end{array}\right).
\ee

\section{Degradation of quantum information under Unruh channel}
\label{sec:degradation}

The  nonclassicality of  quantum information  can be  characterized in
terms   of  nonlocality $B$  (for e.g., Bell inequality violation \cite{CS1978,CHS+69,ADR1982,Per1996,HHH1996,HHH1996/2}),  entanglement, characterized here by concurrence $C$ \cite{Woo1998},
teleportation fidelity $F_{max}$ \cite{BBC+1993,Pop1994,HHH1996} or weaker nonclassicality measures like quantum
discord $D$ \cite{ollivier} or   measurement  induced   disturbance $M$ \cite{luo}.   In  the
accelerated reference frame, the Unruh effect degrades the quantumness
of the  state (\ref{un:matrix}) \cite{us,sbfisher2015}. To achieve our goal, we consider the scenario
wherein only Rob’s qubit is interacting with a noisy environment. The other 
case in which both the qubits of the two observers interact with a noisy environment is not
seen here to produce any qualitatively useful insight and hence is not considered in what follows.

\subsection{Effect of QND noise} 
\label{sec:qnd-degradation}
QND is a purely dephasing noise channel whose action on a qubit, characterized by frequency $\omega_0$, can be studied using the following Kraus operators \cite{sbrg}
\be
K_1=\sqrt{\frac{1-e^{-\omega_0\gamma^2(t)}}{2}}
\left(\begin{array}{ccc}
e^{i\omega_0 t}&0\\
0 & -1
\end{array}\right);~~
K_2=\sqrt{\frac{1+e^{-\omega_0\gamma^2(t)}}{2}}
\left(\begin{array}{ccc}
e^{i\omega_0 t}&0\\
0 & 1
\end{array}\right).
\ee

Assuming an Ohmic bath  spectral density with an upper cut-off frequency $\omega_c$, it can be shown that
\bea
\gamma(t)&=&  \left(\frac{\gamma_0 k_B T}{\pi \hbar \omega_c}\right)
 \cosh(2 s) \left(2 \omega_c t\tan^{-1}(\omega_c t) + 
\ln\left[\frac{1}{1 + \omega_c^2 t^2} \right]\right) \nonumber\\
&&- \left(\frac{\gamma_0 k_B T}{2\pi  \hbar \omega_c}\right) \sinh(2 s) \left(\frac{}{}4 \omega_c(t - a) \tan^{-1}[2 \omega_c (t - a)]
-4 \omega_c (t-2 a) \tan^{-1}[\omega_c (t-2a)] \right.\nonumber\\
&&+ \left. 4 a \omega_c \tan^{-1}(2 a \omega_c) + \ln\left[\frac{\left(1 + \omega_c^2 (t - 2 a)^2\right)^2}{1 + 4\omega_c^2(t-a)^2}\right] + \ln\left[\frac{1}{1 + 4 a^2\omega_c^2}\right]\right).
\eea
Here $T$ is the reservoir temperature, while $a$ and $s$ are bath squeezing parameters. Now, the corresponding Choi matrix have the
form
\be
\left(
\begin{array}{cclr}
\cos^2r&0&0&e^{i\omega_0 t}e^{-\omega_0\gamma^2(t)}\cos r\\
0&\sin^2r&0&0\\
0&0&0&0\\
e^{-i\omega_0 t}e^{-\omega_0\gamma^2(t)}\cos r&0&0&1
\end{array}\right),
\ee

The new Kraus operators are 
\bea
K_1&=&\frac{\sqrt{\cos2 r+3-2 e^{-\frac{1}{4} \left(\gamma  \omega ^2\right)} \sqrt{\mathcal{A}}}}{2 \sqrt{\frac{1}{4} \sec ^2r \left(\sqrt{\mathcal{A}}+e^{\frac{\gamma  \omega ^2}{4}} \sin ^2r\right)^2+1}}
\times
\left(
\begin{array}{cc}
 -\frac{1}{2}e^{i t \omega } \sec r \left( \sqrt{\mathcal{A}}+ e^{\frac{1}{4} \omega^2\gamma }\sin ^2r \right) & 0 \\
 0 & 1 \\
\end{array}
\right)\nonumber\\
\frac{}{}\nonumber\\
K_2&=&\frac{\sqrt{\cos 2 r+3+2 e^{-\frac{1}{4} \left(\gamma  \omega ^2\right)} \sqrt{\mathcal{A}}}}{2 \sqrt{\frac{1}{4} \sec ^2r 
\left(\sqrt{\mathcal{A}}-e^{\frac{\gamma  \omega ^2}{4}} \sin ^2(r)\right)^2+1}}
\times
\left(
\begin{array}{cc}
 \frac{1}{2} e^{i t \omega }\sec r \left( \sqrt{\mathcal{A}}-e^{\frac{1}{4} \omega^2\gamma } \sin ^2r\right) & 0 \\
 0 & 1 \\
\end{array}
\right)\nonumber\\
K_3&=&\left(
\begin{array}{cc}
0&0\\
\sin r&0\\
\end{array}
\right),
\eea
where $\mathcal{A}=e^{\frac{\gamma  \omega ^2}{2}} \sin ^4r + 2 \cos2 r + 2$ and the Kraus operators satisfy the completeness
$\sum_i^3K_i^\dag K_i=\mathbb{I}.$

For the  initial time $t=0$, when  the QND interaction has  not begun,
$e^{-\omega_0\gamma^2(t)}=1$ and  the above Kraus operators  reduce to
that  in Eq.   (\ref{eq:ukraus}).   The composition  of the  dephasing
channel  with the  Unruh channel  has 3  Kraus operators,  essentially
because  only three  of the  resulting four  operators obtained  under
composition, are linearly independent.  To see this, let the two Kraus
operators of the amplitude damping channel be denoted
$$
\mathcal{A}_1 \equiv \left( 
\begin{array}{cc} 1 & 0 \\
0 & \sqrt{1-\lambda}\end{array}\right),\hspace{0.5cm}
\mathcal{A}_2 \equiv \left( \begin{array}{cc} 0 & 0 \\
\sqrt{\lambda} & 0 \end{array}\right),
$$
and those for the dephasing channel by
$$
\mathcal{D}_1 \equiv \sqrt{p}\left( \begin{array}{cc} 1 & 0 \\
0 & 1\end{array}\right),\hspace{0.5cm}
\mathcal{D}_2 \equiv \sqrt{1-p}\left( \begin{array}{cc} 1 & 0 \\
0 & -1\end{array} \right).
$$  The composition  of these  two  channels has  the Kraus  operators
$\mathcal{D}_1\mathcal{A}_1$,            $\mathcal{D}_2\mathcal{A}_1$,
$\mathcal{D}_1\mathcal{A}_2$  and $\mathcal{D}_2\mathcal{A}_2$,  where
the last two terms, namely, 
\begin{equation}
\mathcal{D}_1\mathcal{A}_2 
= \sqrt{p\lambda}\left( \begin{array}{cc} 0
  & 0 \\ 1 & 0 \end{array}\right)~~\textrm{and}~~
\mathcal{D}_2\mathcal{A}_2=\sqrt{p\lambda}\left( \begin{array}{cc} 0
    & 0  \\ -1 &  0 \end{array}\right),
\end{equation}
are equivalent in  that they produce the same noise  effect. Thus, the
composed  noise channel  has  a  rank of  three  corresponding to  the
indepenent      Kraus     operators      $\mathcal{D}_1\mathcal{A}_1$,
$\mathcal{D}_2\mathcal{A}_1$,     and     $\mathcal{D}_1\mathcal{A}_2$
or$\mathcal{D}_2\mathcal{A}_2$.

The QND channel acting on the Unruh qubit effects its quantum characteristics and can be studied by the behavior of the different facets of quantum correlations. For the case of QND noise acting on Rob,
analytical expressions can be obtained for the corresponding measures of quantum correlations which are as follows,
\begin{eqnarray}
M&=&\frac{1}{8}\left[4+\frac{3+\cos2r-2\sqrt{4e^{-\omega_0^2\gamma^4(t)}\cos^2r+\sin^4r}}{8}\log\left(\frac{3+\cos2r-2\sqrt{4e^{-\omega_0^2\gamma^4(t)}\cos^2r+\sin^4r}}{8}  \right)\right.\nonumber\\ 
&&\left.  +\frac{3+\cos2r+2\sqrt{4e^{-\omega_0^2\gamma^4(t)}\cos^2r+\sin^4r}}{8}\log\left(\frac{3+\cos2r+2\sqrt{4e^{-\omega_0^2\gamma^4(t)}\cos^2r+\sin^4r}}{8}  \right) 
\right.\nonumber\\ 
&&\left.-4\cos^2r\log\left(\frac{\cos^2r}{2}\right) \right],
\end{eqnarray}

\begin{equation}
F_{\rm{max}}=\frac{1}{2}\left[1+\frac{\cos r}{3}\left(2e^{-\omega_0\gamma^2(t)}+\cos r\right)\right],
\end{equation}

\begin{equation}
B=2e^{-\omega_0^2\gamma^4(t)}\cos^2r.
\end{equation}

The analytical expression for entanglement $C$ turns out to be very involved. Hence we only provide a numerical analysis.
For the initial time $t=0$ when QND interaction has not begun, $e^{-\omega_0\gamma^2(t)}=1$ and
the expressions $F_{max}$, $B$ and $M$ in the above equations reduce to the pure Unruh-effect case.

\begin{figure}
\begin{center}
\subfigure[]{
\includegraphics[width=0.35\textwidth]{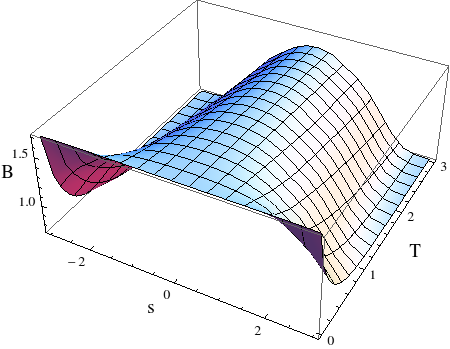}}
\subfigure[]{
\includegraphics[width=0.35\textwidth]{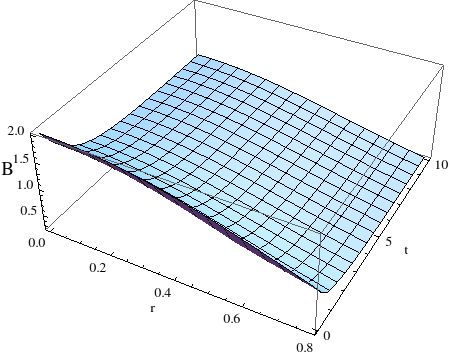}}
\end{center}
\caption{(a) Plot of Bell's inequality $B$ against temperature $T$ and squeezing $s$ 
when  Robs accelerated qubit interacts dissipatively with
the environment.  Robs acceleration
corresponds to $r=\pi/8$ and $t=0.5$. (b)  Plot of Bell's inequality $B$ against time $t$ for which QND channel acts and $r$.     
The bath parameters are $\omega_0=1$, the coupling $\gamma=0.1$ and squeezing angle
$\phi=0$, $T=0.5$ and $s=0.5$.\label{bell-qnd} }
\end{figure}

\begin{figure}
\begin{center}
\subfigure[]{
\includegraphics[width=0.35\textwidth]{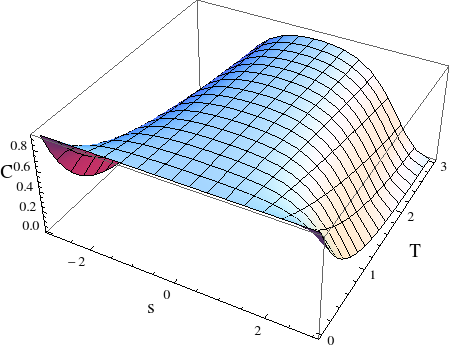}}
\subfigure[]{
\includegraphics[width=0.35\textwidth]{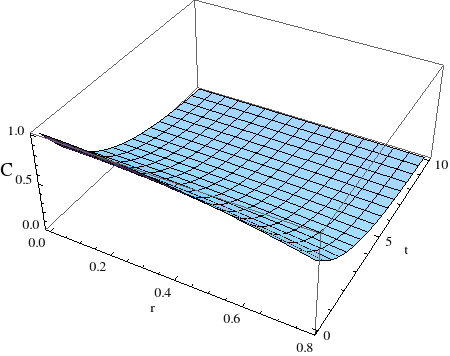}}
\end{center}
\caption{(a) Plot of concurrence $C$ against temperature $T$ and squeezing $s$ 
when  Robs accelerated qubit is subjected to a QND channel.  Robs acceleration
corresponds to $r=\pi/8$ and $t=0.5$. (b)  Plot of  $C$ against time $t$ for which QND channel acts and $r$.     
The bath parameters are $\omega_0=0.1$, the coupling $\gamma=0.1$ and squeezing angle
$\phi=0$, $T=1.5$ and $s=1.5$.\label{concu-qnd}}
\end{figure}

\begin{figure}
\begin{center}
\subfigure[]{
\includegraphics[width=0.35\textwidth]{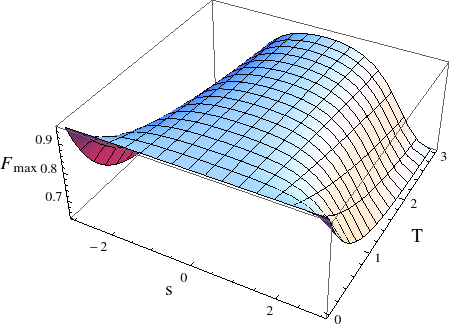}}
\subfigure[]{
\includegraphics[width=0.35\textwidth]{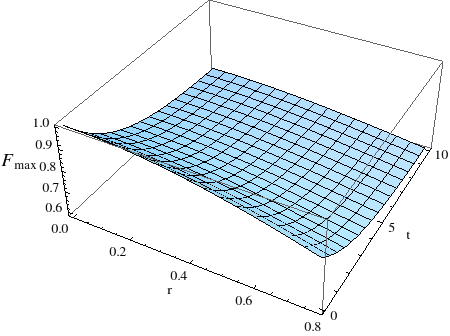}}
\end{center}
\caption{(a) Plot of teleportation fidelity $F_{\rm max}$ against temperature $T$ and squeezing $s$ 
when  Robs accelerated qubit is subjected to a QND channel.  Robs acceleration
corresponds to $r=\pi/8$, $t=0.5$. (b)  Plot of $F_{\rm max}$ against time $t$ for which QND channel acts and $r$.     
The bath parameters are $\omega_0=1$, the coupling $\gamma=0.1$ and squeezing angle
$\phi=0$, $T=1.5$ and $s=1.5$.\label{teleport-qnd}}
\end{figure}

\begin{figure}
\begin{center}
\subfigure{
\includegraphics[width=0.35\textwidth]{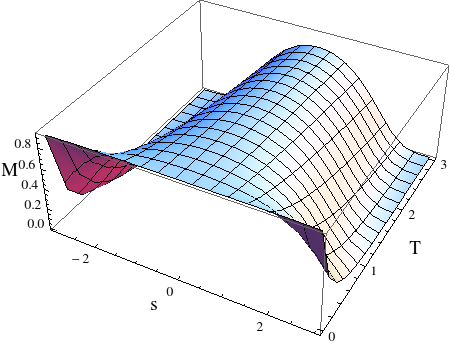}}
\subfigure{
\includegraphics[width=0.35\textwidth]{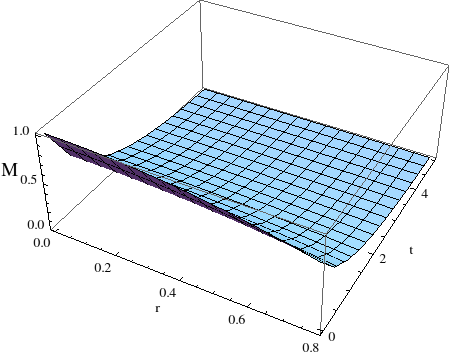}}
\end{center}
\caption{(a) Plot of $M$ against temperature $T$ and squeezing $s$ 
when  Robs accelerated qubit is subjected to a QND channel.  Robs acceleration
corresponds to $r=\pi/8$ and $t=0.5$. (b)  Plot of $M$ against time $t$ for which QND channel acts and $r$.     
The bath parameters are $\omega_0=1$, the coupling $\gamma=0.1$ and squeezing angle
$\phi=0$, $T=1.5$ and $s=1.5$.\label{m-qnd}}
\end{figure}

The figures \ref{bell-qnd} to \ref{m-qnd} correspond to the behavior of various aspects of quantum
correlations of the Unruh channel under the influence of QND noise. 
It can be seen from Fig.~\ref{bell-qnd}(a) that as reservoir squeezing $s$ increases, the channel becomes local
even for a small external temperature $T$. Also, from Fig.~\ref{bell-qnd}(b) the channel is seen to become local
with increase in the Unruh acceleration depicted, here, by $r$. Entanglement is seen, in Fig.~\ref{concu-qnd}(a), 
to decrease with increase in $s$. This feature is more prominent for $T>1$. From  Fig.~\ref{concu-qnd}(b), for a 
given value of $r$, entanglement is seen to decrease with time. For $|s|<2$, Fig.~\ref{teleport-qnd}(a) shows that
$F_{\rm max}>\frac{2}{3}$  for the given temperature range. Also, $F_{\rm max}$ decreases with increase in $r$ 
and time of evolution $t$, Fig.~\ref{teleport-qnd}(b).
$M$,  Fig.~\ref{m-qnd}, is seen to decrease with increase in the parameters $t$, $s$, $T$ and $r$.

\subsection{Effect of SGAD noise}
\label{sec:sgad-degradation}

The Kraus corresponding to the SGAD channel are 
\begin{eqnarray}
\begin{array}{ll}
K_{1} \equiv \sqrt{p_1}\left[\begin{array}{ll} 
\sqrt{1-\alpha} & 0 \\ 0 & 1
\end{array}\right], ~~~~ &
K_{2} \equiv \sqrt{p_1}\left[\begin{array}{ll} 0 & 0 \\ \sqrt{\alpha} & 0
\end{array}\right],  \\
K_{3} \equiv \sqrt{p_2}\left[\begin{array}{ll} \sqrt{1-\mu} & 0 \\ 0 & 
\sqrt{1-\nu}
\end{array}\right], ~~~~ &
K_{4} \equiv \sqrt{p_2}\left[\begin{array}{ll} 0 & \sqrt{\nu} \\ \sqrt{\mu}e^{-i\phi_s} & 0
\end{array}\right],
\end{array}
\label{eq:srikraus}
\end{eqnarray}
where $p_1+p_2=1$ \cite{sbsrik}. Here 
\begin{eqnarray}
p_2 &=& \frac{1}{(A+B-C-1)^2-4D}
 \times \left[A^{ 2} B + C^{2} + A(B^{2} - C - B(1+C)-D) - (1+B)D 
\nonumber \right. \\
&&- C(B+D-1) \pm \left.  2\sqrt{D(B-A B+(A-1)C+D)(A-A B+(B-1)C+D)}\right],
\label{eq:p2}
\end{eqnarray}
where
\begin{eqnarray}
A &=& \frac{2N+1}{2N} \frac{\sinh^2(\gamma_0 at/2)}
{\sinh(\gamma_0(2N+1)t/2)}
\exp\left(-\gamma_0(2N+1)t/2\right),~~B = \frac{N}{2N+1}(1-\exp(-\gamma_0(2N+1)t)), \nonumber \\
C &= & A + B + \exp(-\gamma_0 (2N+1)t),~~
D = \cosh^2(\gamma_0 at/2)\exp(-\gamma_0(2N+1)t).
\label{eq:auxip2}
\end{eqnarray}
Also,
\begin{eqnarray}
\nu &=& \frac{N}{(p_2)(2N+1)}(1-e^{-\gamma_0(2N+1)t}),~~
\mu = \frac{2N+1}{2(p_2) N}\frac{\sinh^2(\gamma_0at/2)}{\sinh(\gamma_0(2N+1)t/2)}
\exp\left(-\frac{\gamma_0}{2}(2N+1)t\right),\nonumber \\
\alpha & =& \frac{1}{p_1}\left(1 - p_2[\mu(t)+\nu(t)]
- e^{-\gamma_0(2N+1)t}\right).
\label{eq:nu}
\end{eqnarray} 
Also 
$ N = N_{\rm th}[\cosh^2(s) + \sinh^2(s)]  + \sinh^2(s),~~a=\sinh(2s)( 2N_{\rm th}+1)$ where 
$N_{\rm th}= 1/(e^{\hbar  \omega_0/k_B T} -  1)$ is the Planck  distribution giving
the  number of  thermal photons  at  the frequency  $\omega_0$; $s$  and
$\phi_s$  are  bath  squeezing  parameters.

\begin{figure}
\begin{center}
\subfigure[]{
\includegraphics[width=0.35\textwidth]{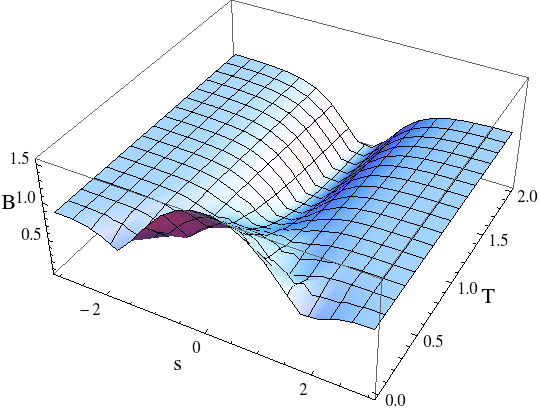}}
\subfigure[]{
\includegraphics[width=0.35\textwidth]{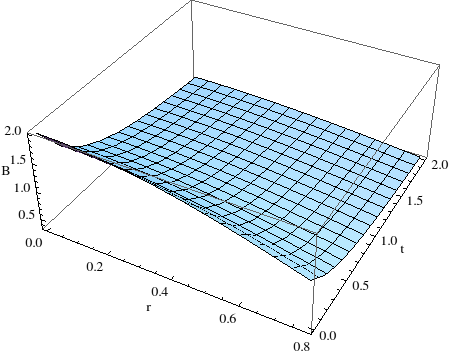}}
\end{center}
\caption{(a) Plot of Bell's inequality $B$ against temperature $T$ and squeezing $s$ 
when  Robs accelerated qubit is subjected to a SGAD channel.  Robs acceleration
corresponds to $r=\pi/8$. (b)  Plot of Bell's inequality $B$ against time $t$ for which SGAD channel acts and $r$.     
The bath parameters are $\omega_0=0.1$, the coupling $\gamma=0.1$ and squeezing angle
$\phi=0$, $T=0.5$ and $s=0.5$. \label{bell-sgad} }
\end{figure}

\begin{figure}
\begin{center}
\subfigure[]{
\includegraphics[width=0.35\textwidth]{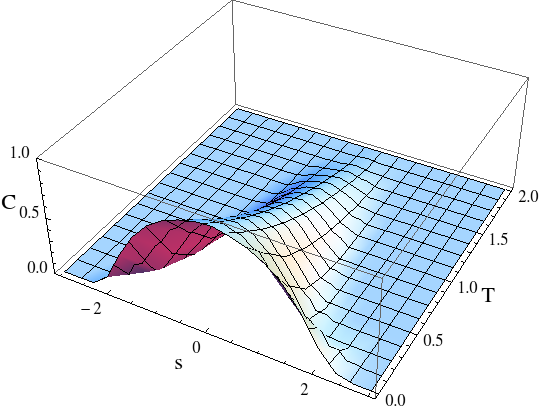}}
\subfigure[]{
\includegraphics[width=0.35\textwidth]{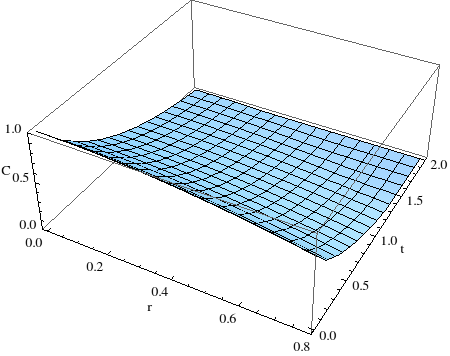}}
\end{center}
\caption{(a) Plot of concurrence $C$ against temperature $T$ and squeezing $s$ 
when   Robs accelerated qubit is subjected to a SGAD channel.  Robs acceleration
corresponds to $r=\pi/8$. (b)  Plot of  $C$ against time $t$ for which SGAD channel acts and $r$.     
The bath parameters are $\omega_0=0.1$, the coupling $\gamma=0.1$ and squeezing angle
$\phi=0$, $T=0.5$ and $s=0.5$.\label{concu-sgad}}
\end{figure}

\begin{figure}
\begin{center}
\subfigure[]{
\includegraphics[width=0.35\textwidth]{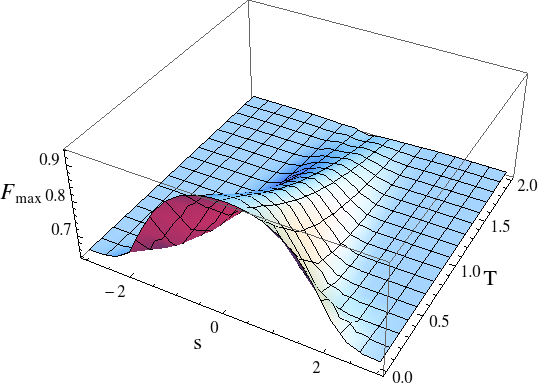}}
\subfigure[]{
\includegraphics[width=0.35\textwidth]{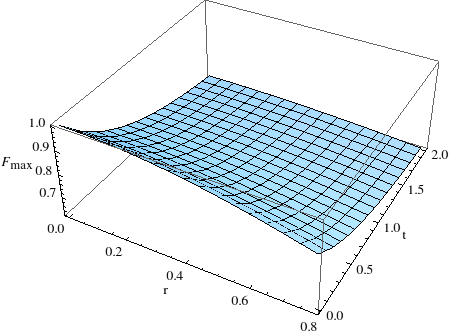}}
\end{center}
\caption{(a) Plot of teleportation fidelity $F_{\rm max}$ against temperature $T$ and squeezing $s$ 
when the  Robs accelerated qubit is subjected to a SGAD channel.  Robs acceleration
corresponds to $r=\pi/8$. (b)  Plot of $F_{\rm max}$ against time $t$ for which SGAD channel acts and $r$.     
The bath parameters are $\omega_0=0.1$, the coupling $\gamma=0.1$ and squeezing angle
$\phi=0$, $T=0.5$ and $s=0.5$.\label{tele-sgad}}
\end{figure}

\begin{figure}
\begin{center}
\subfigure{
\includegraphics[width=0.35\textwidth]{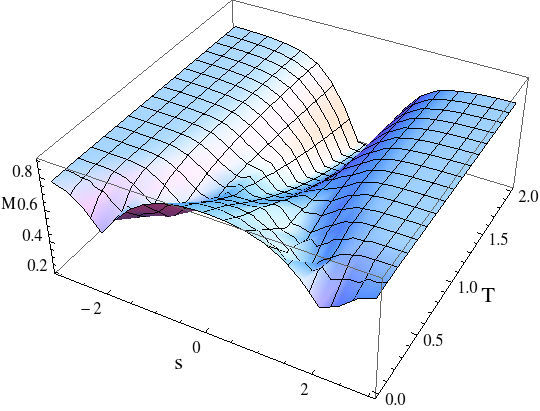}}
\subfigure{
\includegraphics[width=0.35\textwidth]{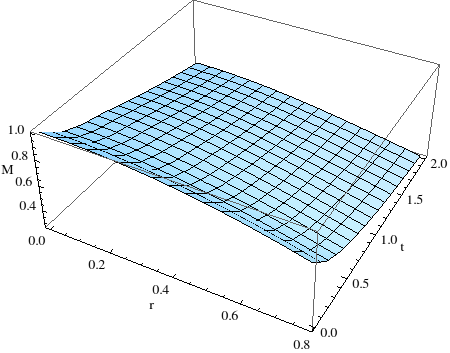}}
\end{center}
\caption{(a) Plot of MID $M$ against temperature $T$ and squeezing $s$ 
when the  Robs accelerated qubit is subjected to a SGAD channel.  Robs acceleration
corresponds to $r=\pi/8$ and $t=0.5$. (b)  Plot of $M$ against time $t$ for which SGAD channel acts and $r$.     
The bath parameters are $\omega_0=0.1$, the coupling $\gamma=0.1$ and squeezing angle
$\phi=0$, $T=0.5$ and $s=0.5$.\label{m-sgad}}
\end{figure}

The analytical expressions are complicated and hence we resort to numerical discussions.
The figures \ref{bell-sgad} - \ref{m-sgad} correspond to the behavior of various aspects of quantum
correlations of the Unruh channel under the influence of SGAD noise. From Fig.~\ref{bell-sgad}(a) it can be seen that
for certain range of $T$, squeezing enhances $B$. However, for the values of the parameters indicated, it never crosses
the threshold of nonlocality ($B>1$). From Fig.~\ref{bell-sgad}(b) it can be seen that with increase in $r$ and $t$, $B$ decreases 
and the channel becomes local. 
From Fig.~\ref{concu-sgad}(a), concurrence
is seen to drop drastically to zero with increase in $T$. Also for large values of $s$,  concurrence is seen to
fall to zero, irrespective of $T$. Fig.\ref{concu-sgad}(b) depicts the decrease of concurrence with respect to time
for any give value of $r$. From Fig.~\ref{tele-sgad}, $F_{\rm max}$ is seen to decrease with $T$, $s$, $r$ and $t$. 
From  Fig.~\ref{m-sgad}, $M$ is seen to decrease with increase in the parameters $r$ and $t$. However for certain range of $T$, $M$ 
is seen to increase with bath squeezing $s$, reiterating the usefulness of squeezing.

\section{Average gate and Channel fidelity}
In this work we are trying to understand the Unruh channel under the influence of external noisy effects. 
From the perspective of quantum information, a useful way to understand this is to analyze the average gate and 
channel fidelities \cite{gate} of the Unruh channel under the ambient environmental conditions. 

The average gate fidelity \cite{sbsrik,gate1,gate2,gate3} has a closed expression
\be
G_{{\rm av}}=\frac{d+\sum_i|{\rm Tr}(E_i)|^2}{d(d+1)}.
\ee
For the Unruh channel $G_{{\rm av}}=\frac{1}{6}\left(2+(1+\cos r)^2\right)$, where $d$ is the dimensionality of the system
on which channel $\mathcal{E}$acts with operator sum representation elements $E_i$. 

For the QND-Unruh channel
\bea
|\rm{Tr}(E_i)|^2=\frac{\left(2 \sqrt{\mathcal{A}} e^{-\frac{1}{4} \gamma  \omega_0 ^2}+\cos 2r+3\right) \left(\sec r \cos \omega_0 t \left(\sqrt{\mathcal{A}}-e^{\frac{\gamma  \omega_0 ^2}{4}} \sin ^2r\right)+\frac{1}{4} \sec ^2r \left(\sqrt{\mathcal{A}}-e^{\frac{\gamma  \omega_0 ^2}{4}} \sin ^2r\right)^2+1\right)}{\sec ^2r \left(\sqrt{\mathcal{A}}-e^{\frac{\gamma  \omega_0 ^2}{4}} \sin ^2r\right)^2+4}\nonumber\\
+\frac{\left(-2 \sqrt{\mathcal{A}} e^{-\frac{1}{4} \gamma  \omega_0 ^2}+\cos r+3\right) \left(-\sec r \cos \omega_0 t \left(\sqrt{\mathcal{A}}+e^{\frac{\gamma  \omega_0 ^2}{4}} \sin ^2r\right)+\frac{1}{4} \sec ^2r \left(\sqrt{\mathcal{A}}+e^{\frac{\gamma  \omega_0 ^2}{4}} \sin ^2r\right)^2+1\right)}{\sec ^2r \left(\sqrt{\mathcal{A}}+e^{\frac{\gamma  \omega_0 ^2}{4}} \sin ^2r\right)^2+4},
\eea
using which $G_{\rm{av}}$ can be calculated. In the limit $t\longrightarrow0$ this reduces to the Unruh case. 
Unlike the QND case, the analytical expression for $G_{{\rm av}}$ for the SGAD channel is very involved, and hence we discuss this case
numerically.

\begin{figure}
\begin{center}
\subfigure{
\includegraphics[width=0.35\textwidth]{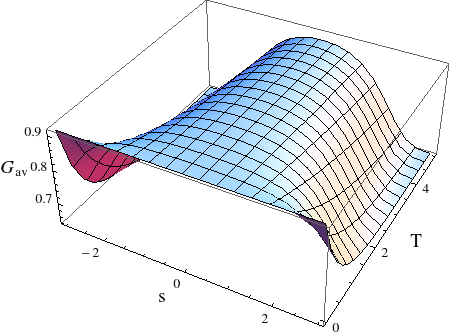}}
\subfigure{
\includegraphics[width=0.35\textwidth]{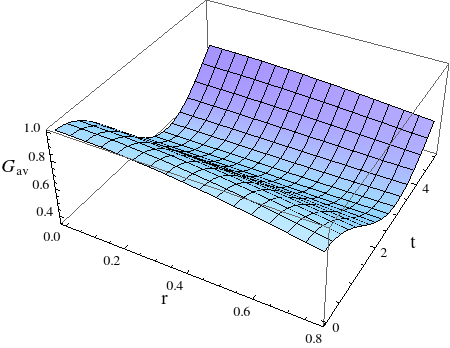}}
\end{center}
\caption{(a) Plot of average gate fidelity $G_{{\rm av}}$ against temperature $T$ and squeezing $s$ 
when   Robs accelerated qubit is subjected to a QND channel.  Robs acceleration
corresponds to $r=\pi/8$ and $t=0.5$. (b)  Plot of  $G_{{\rm av}}$ against time $t$ for which QND channel acts and $r$.     
The bath parameters are $\omega_0=1$, the coupling $\gamma=0.1$ and squeezing angle
$\phi=0$, $T=0.5$ and $s=0.5$.\label{gav-qnd}}
\end{figure}

\begin{figure}
\begin{center}
\subfigure{
\includegraphics[width=0.35\textwidth]{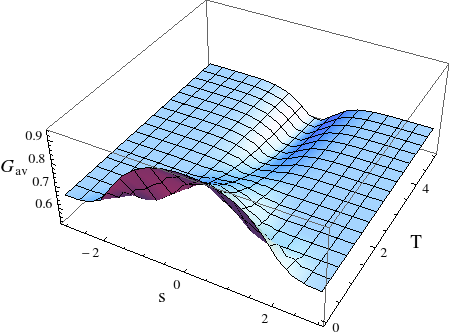}}
\subfigure{
\includegraphics[width=0.35\textwidth]{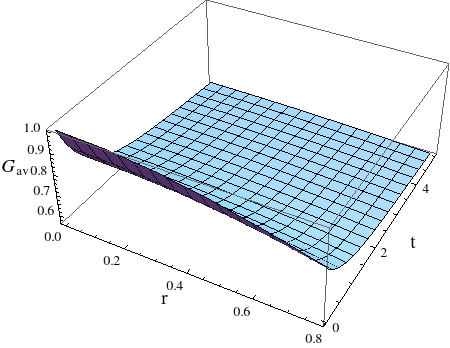}}
\end{center}
\caption{(a) Plot of average gate fidelity $G_{{\rm av}}$ against temperature $T$ and squeezing $s$ 
when   Robs accelerated qubit is subjected to a SGAD channel.  Robs acceleration
corresponds to $r=\pi/8$ and $t=0.5$. (b)  Plot of $G_{{\rm av}}$ against time $t$ for which SGAD channel acts and $r$.     
The bath parameters are $\omega_0=0.1$, the coupling $\gamma=0.1$ and squeezing angle
$\phi=0$, $T=0.5$ and $s=0.5$. \label{gav-sgad}}
\end{figure}

From Fig.~\ref{gav-qnd}(a), it can be seen that $G_{{\rm av}}$ is stable for a certain range of squeezing, after which it falls. $G_{{\rm av}}$ is also seen to decrease with $r$. However with time, 
$G_{{\rm av}}$ first decreases and then is then seen to increase.  
 Since the expression for $G_{\rm{av}}$ has the oscillatory term $\cos\omega_0 t$,  oscillation is seen in Fig. \ref{gav-qnd}(b) with time $t$. 

From Figs.~\ref{gav-sgad}, a general trend is observed of average gate fidelity $G_{\rm av}$, under the influence of the SGAD channel, decreasing with increase in $T$
as well as evolution time $t$, for a given $r$. However, as can be observed from Fig.~\ref{gav-sgad}(a), for a certain range of $T$,
$G_{\rm av}$ is seen to increase with reservoir squeezing $s$. This indicates that squeezing can be a useful quantum resource in this
scenario.

Another quantity frequently used to access the channel's performance is the channel fidelity \cite{sbsrik,gate}
\be
\chi=\max_{\mathcal{B}}\kappa(\mathcal{B},\mathcal{E}),
\ee
where $\kappa(\mathcal{B},\mathcal{E})$ is the Holevo bound for states prepared in basis $\mathcal{B}$ and passed through 
the channel $\mathcal{E}$. The maximum is achieved for the states prepared in the basis $\{|0\rangle,|1\rangle\}$, i.e., for the state $\frac{1}{2}( |0\rangle\langle0|+ |1\rangle \langle1|)$. 
 $\kappa$ signifies how well the quantum input states are distinguishable after the action of the channel $\mathcal{E}$. It can also be interpreted as how much classical information can be extracted from  the given quantum ensemble.

\begin{figure}
\begin{center}
\subfigure{
\includegraphics[width=0.35\textwidth]{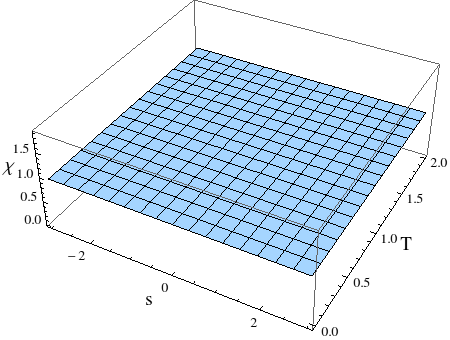}}
\subfigure{
\includegraphics[width=0.35\textwidth]{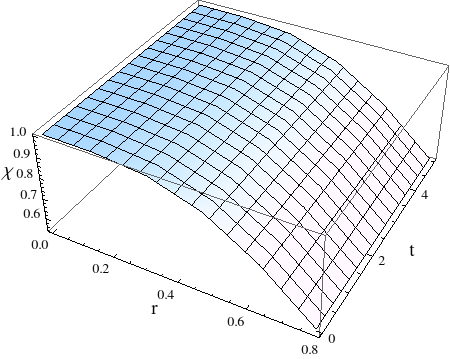}}
\end{center}
\caption{(a) Plot of channel fidelity $\chi$ against temperature $T$ and squeezing $s$ 
when  Robs accelerated qubit is subjected to a QND channel.  Robs acceleration
corresponds to $r=\pi/8$ and $t=0.5$. (b)  Plot of  $\chi$ against time $t$ for which QND channel acts and $r$.     
The bath parameters are $\omega_0=1$, the coupling $\gamma=0.1$ and squeezing angle
$\phi=0$, $T=0.5$ and $s=0.5$.\label{chi-qnd}}
\end{figure}

\begin{figure}
\begin{center}
\subfigure{
\includegraphics[width=0.35\textwidth]{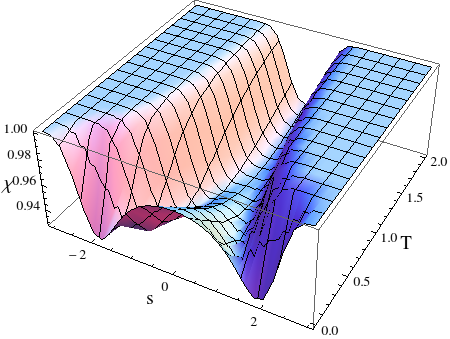}}
\subfigure{
\includegraphics[width=0.35\textwidth]{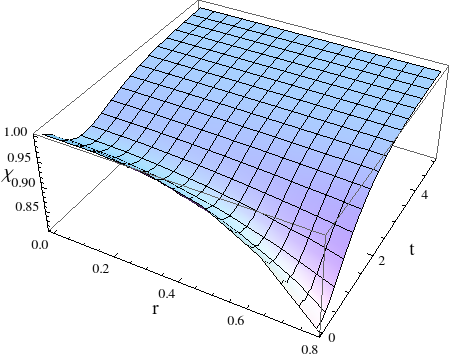}}
\end{center}
\caption{(a) Plot of channel fidelity $\chi$ against temperature $T$ and squeezing $s$ 
when  Robs accelerated qubit is subjected to a SGAD channel.  Robs acceleration
corresponds to $r=\pi/8$ and $t=0.5$. (b)  Plot of $\chi$ against time $t$ for which SGAD channel acts and $r$.     
The bath parameters are $\omega_0=0.1$, the coupling $\gamma=0.1$ and squeezing angle
$\phi=0$, $T=0.5$ and $s=0.5$.\label{chi-sgad}}
\end{figure}

From Fig.~\ref{chi-qnd} it can be seen that $\chi$, for the Unruh-QND channel, does not change with $s$, $T$ and $t$.  The reason is that the maximum is obtained for completely mixed state with no off-diagonal terms which make the QND ineffective, whereas the diagonal terms are altered by the Unruh effect. 
For the Unruh-SGAD channel, we see from Fig.~\ref{chi-sgad}(a)
that the Unruh channel drives the state away from the maximally mixed state for which the Holevo bound is maximum. By increasing the bath squeezing and the temperature the system is driven back towards the maximally mixed states which results in increment of  $\chi$ to its  maximum value. Fig.~\ref{chi-sgad}(b) show that as the SGAD channel acts for longer duration the state is driven towards maximally mixed state and thus $\chi$ increases with  $t$.

\section{Coherence and Mixedness}
 Quantum coherence  as a resource has attracted much attention in recent times  \cite{coh1,coh3,coh4}.
Coherence plays a central role in quantum mechanics \cite{Alok:2014gya} enabling operations or tasks which are impossible within the regime of classical mechanics. 

For a quantum state represented by density matrix $\rho$ in basis $\{|i\rangle\}$, the $l_1$ norm of coherence is given by
\be
{\cal C}(\rho)=\sum_{i\neq j}|\rho_{i,j}|.
\ee

The mixedness, which is basically normalized
linear entropy,  of a $d$ dimensional quantum state $\rho$ is given by \cite{mix}
\be
{\cal M}(\rho)=\frac{d}{d-1}(1-{\rm Tr}\rho^2).
\ee

The inequality relation between the ${\cal C}(\rho)$ and ${\cal M}(\rho)$ is \cite{coh4}
\be
\frac{{\cal C}(\rho)^2}{(d-1)^2}+{\cal M}(\rho)\leq1.
\label{eq:inq}
\ee
When this inequality is saturated for certain values of ${\cal C}(\rho)$ and ${\cal M}(\rho)$, the situation corresponds to 
states which have maximum coherence for a given mixedness in the states. Such states are known as Maximally Coherent Mixed States (MCMS).

\subsection{QND Channel}
The analytical expressions for coherence and mixedness of the Unruh channel under the influence of the QND noise are
\be
{\cal C}(\rho)=\cos\frac{\theta}{2}\sin\frac{\theta}{2}\cos^r e^{-\frac{\omega_0^2\gamma(t)}{4}},
\ee

\be
{\cal M}(\rho)=\cos^2r\left(\cos^2\frac{\theta}{2}(3-\cos2r-2\cos^2r\cos\theta)- e^{-\frac{\omega_0^2\gamma(t)}{2}}\sin^2\theta \right),
\ee
respectively.

\begin{figure}
\begin{center}
\subfigure[]{
\includegraphics[width=0.35\textwidth]{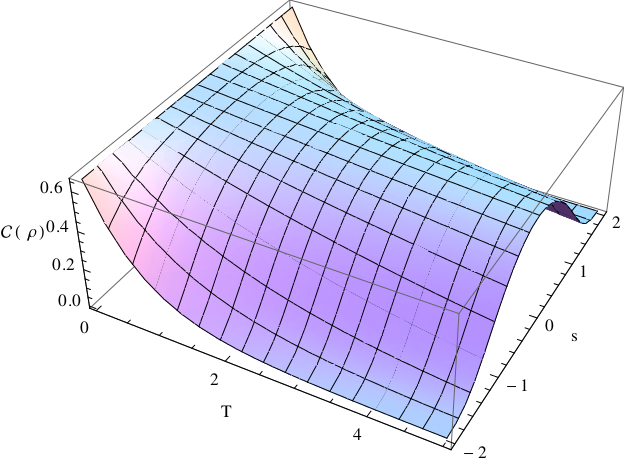}}
\subfigure[]{
\includegraphics[width=0.35\textwidth]{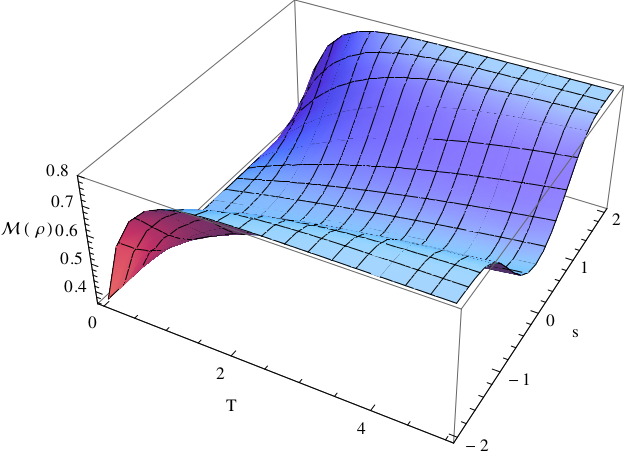}}
\end{center}
\caption{ (a)Variation of  Coherence due to action of QND channel with bath squeezing (s) and temperature (T). (b) Variation of Mixedness plotted against s and T. The bath parameters are $a=0$, $\omega_0=1$, $\gamma=0.1$, $t=2$, $r=\pi/8$ and the input state is parameterized by   $\theta=\pi/4$, $\phi=\pi/4$.\label{cm-qnd} }
\end{figure}

\begin{figure}
\begin{center}
\subfigure[]{
\includegraphics[width=0.35\textwidth]{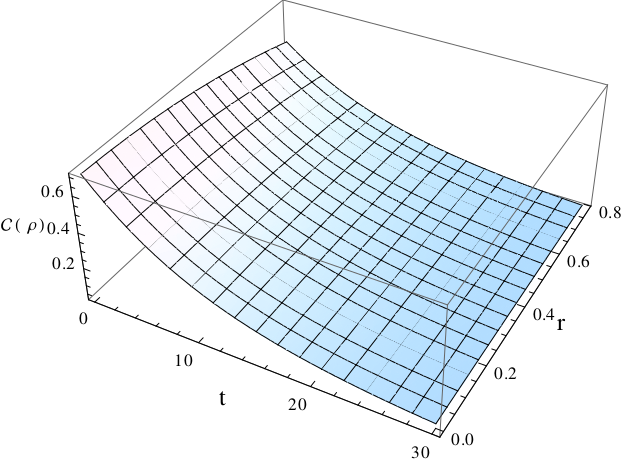}}
\subfigure[]{
\includegraphics[width=0.35\textwidth]{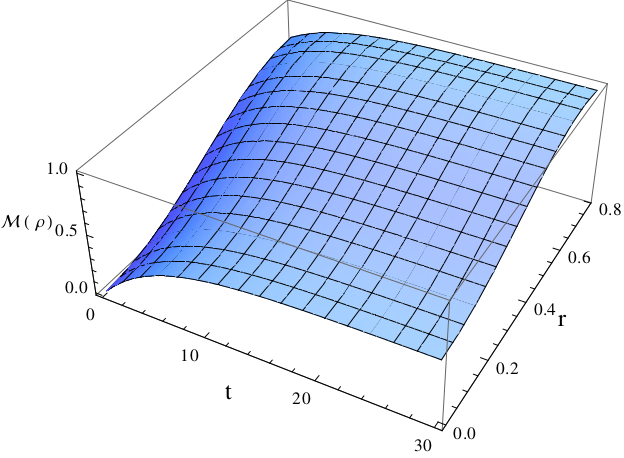}}
\end{center}
\caption{Action of QND channel on (a) Coherence (b) Mixedness plotted against time (t) and Unruh parameter (r). 
The bath parameters are $a=0$, $\omega_0=1$, $\gamma=0.5$, $T=0.5$, $s=0.5$ and the input state is parameterized by $\theta=\pi/4$, $\phi=\pi/4$.\label{cm-qnd-tr}}
\end{figure}

From Fig.~\ref{cm-qnd}, coherence ${\cal C}$ is seen to decrease with increase in temperature $T$ as well as reservoir squeezing $s$. 
Mixing ${\cal M}$ increases with both $T$ and $s$. From Fig.~\ref{cm-qnd-tr}, it can be seen that coherence decreases with $t$ whereas mixedness increases with $t$ as well as $r$, a feature which is
consistent with common intuition.

\subsection{SGAD Channel}
The analytical expressions for coherence and mixedness of the Unruh channel under the influence of the SGAD noise is
\be
{\cal C}(\rho)=\cos r\sin\theta\sqrt{(p_1\sqrt{1-\alpha}+ p_2\sqrt{(1-\mu)(1-\nu)})^2 +p_2^2\mu\nu+2\cos(2\phi-\phi_s)\sqrt{\mu\nu}(p_1\sqrt{1-\alpha}+ p_2\sqrt{(1-\mu)(1-\nu)}) },
\ee

\bea
{\cal M}(\rho)&=& 2 \left(1 -\left(\cos\frac{\theta}{2}^2 ((p_1 + p_2 - p_1 \alpha - p_2 \mu) \cos^2r + p_2 \nu \sin^2r) + p_2 \nu \sin^2\frac{\theta}{2}\right)^2\right.\nonumber\\
&& - \left((p_1 \alpha + p_2 \mu) \cos^2r \cos^2\frac{\theta}{2} + (p_1 + p_2 - p_2 \nu) (\cos^2\frac{\theta}{2}\sin^2r +
\sin^2\frac{\theta}{2})\right)^2 \nonumber\\
&&- \frac{1}{2} e^{-i \phi - i (\phi + \phi_s)} \left(e^{i \phi_s} (p_1 \sqrt{1 - \alpha} + p_2 \sqrt{1 - \mu} \sqrt{1 - \nu}) + e^{2 i \phi} p_2 \sqrt{\mu}  \sqrt{\nu}\right) \times\nonumber\\
&&\left.\left( e^{2 i \phi} (p_1 \sqrt{1 - \alpha} + p_2 \sqrt{1 - \mu} \sqrt{1 - \nu}) + e^{i \phi_s} p_2 \sqrt{\mu} \sqrt{\nu}\right) \cos^2r \sin^2\theta\right),
\eea

\begin{figure}
\begin{center}
\subfigure[]{
\includegraphics[width=0.35\textwidth]{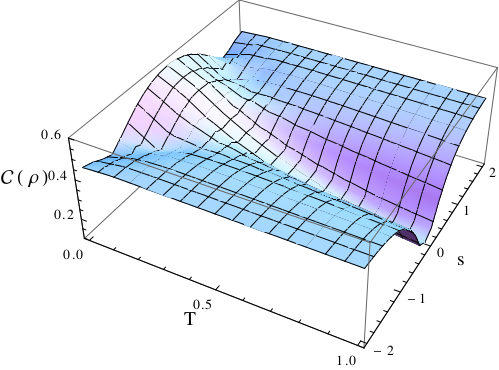}}
\subfigure[]{
\includegraphics[width=0.35\textwidth]{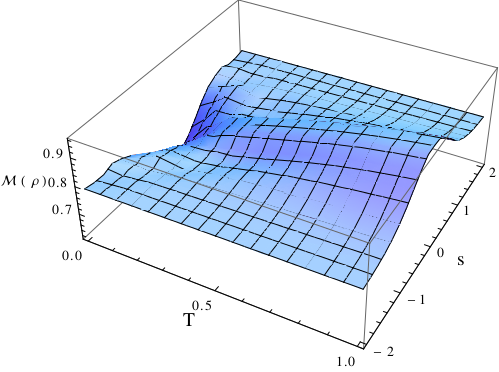}}
\end{center}
\caption{Action of SGAD channel on (a) Coherence (b) Mixedness plotted
  against bath squeezing (s)  and temperature(T).  The bath parameters
  are  $\phi_s=0$, $\omega_0=.1$,  $\gamma=0.1$, $t=2$,  $r=\pi/8$ and
  the    input    state    is   parameterized    by    $\theta=\pi/4$,
  $\phi=\pi/4$.\label{cm-sgad}}
\end{figure}

\begin{figure}
\begin{center}
\subfigure[]{
\includegraphics[width=0.35\textwidth]{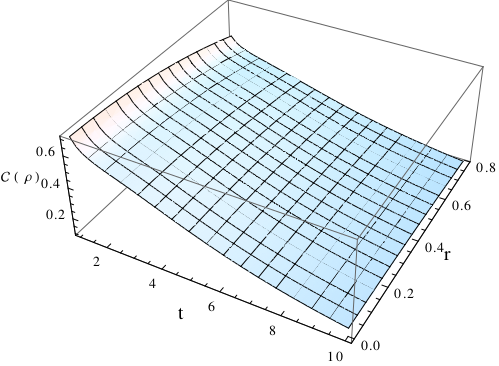}}
\subfigure[]{
\includegraphics[width=0.35\textwidth]{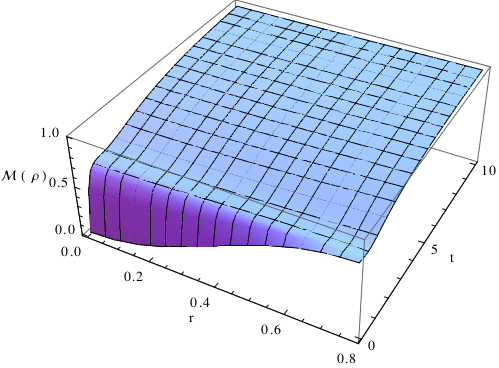}}
\end{center}
\caption{(a)  Variation of  Coherence due  to action  of SGAD  channel
  w.r.t  time (t)  and  Unruh  parameter (r).  (b)  Plot of  Mixedness
  againsttime (t) and Unruh parameter (r). The bath parameters are $\phi_s=0$, $\omega_0=.1$, $\gamma=0.1$, $T=0.5$, $s=0.5$ and the input state is parameterized by $\theta=\pi/4$, $\phi=\pi/4$. \label{cm-sgad-tr}}
\end{figure}

respectively. From Fig.~\ref{cm-sgad}(a), it is observed that for a certain range of temperature $T$, coherence ${\cal C}$ increases with squeezing $s$ while in another range, it decreases with $s$, 
in consistence with the quadrature nature of squeezing. Also, finite $s$ brings a notion of stability in the behavior of coherence ${\cal C}$ as a function of external temperature $T$.
 Thus, squeezing $s$ once again emerges as a useful resource. The expected increase in mixing ${\cal M}$ with increase in $T$ is observed in Fig.~\ref{cm-sgad}(b). 
 From Fig.~\ref{cm-sgad-tr}, it can be seen that coherence decreases with increase in $t$ and $r$ whereas mixedness increases rapidly with $t$.

\begin{figure}
\begin{center}
\subfigure[]{
\includegraphics[width=0.35\textwidth]{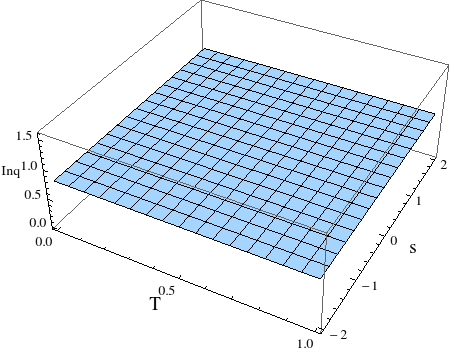}
\label{fig:inq-qnd-s-tem}}
\subfigure[]{
\includegraphics[width=0.35\textwidth]{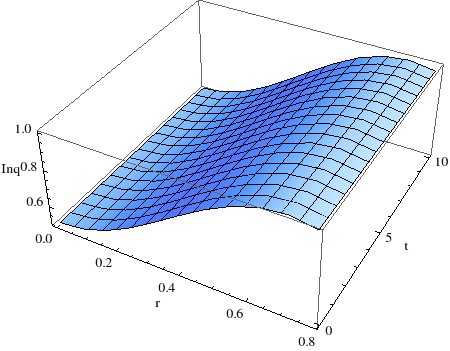}
\label{fig:inq-qnd-t-r}}
\end{center}
\caption{Plots of inequality (Inq) in Eq. (\ref{eq:inq}) due to action of QND (a) w.r.t parameters $T$ and $s$ with $t=2$ and $r=\pi/8$, (b) against time (t) and Unruh parameter (r) with $T=0.5$, $s=0.5$. The bath parameters are $\phi_s=0$, $\omega_0=0.1$, $\gamma=0.1$ and the input state parameterized by $\theta=\pi/4$, $\phi=\pi/4$. }
\label{fig:inq-qnd}
\end{figure}

\begin{figure}
\begin{center}
\subfigure[]{
\includegraphics[width=0.35\textwidth]{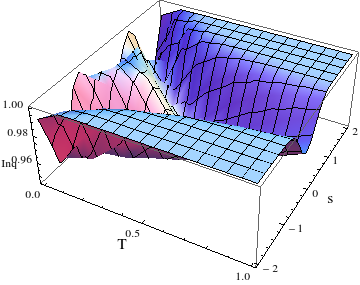}
\label{fig:inq-sgad-s-tem}}
\subfigure[]{
\includegraphics[width=0.35\textwidth]{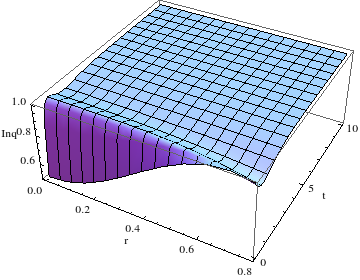}
\label{fig:inq-sgad-t-r}}
\end{center}
\caption{Plot of inequality (Inq) in Eq. (\ref{eq:inq}) due to action of SGAD channel (a) w.r.t parameters $T$ and $s$ with $t=2$ and $r=\pi/8$, (b)  against time (t) and Unruh parameter (r) with $T=0.5$, $s=0.5$. The bath parameters are $\phi_s=0$, $\omega_0=0.1$, $\gamma=0.1$ and the input state parameterized by $\theta=\pi/4$, $\phi=\pi/4$.}
\label{fig:inq-sgad}
\end{figure}

From Figs.   \ref{fig:inq-qnd} and \ref{fig:inq-sgad}, it  is clear
that the inequality   Eq. (\ref{eq:inq}) is respected for both QND and SGAD noises, respectively. In particular, from Fig. \ref{fig:inq-qnd-s-tem}, due to the action of QND noise, it can be seen that for $t=2$ and $r=\pi/8$, by varying parameters $T$ and $s$ we cannot get any MCMS as the inequality is not saturated. However, fixing  $T$ and $s$ to 0.5 and varying  $r$, MCMS can be achieved, as can be seen from Fig. \ref{fig:inq-qnd-t-r}. For the case of SGAD noise, Fig. \ref{fig:inq-sgad-s-tem} depicts that for $t=2$ and $r=\pi/8$, by varying the parameters $T$ and $s$  MCMS states can be attained. Also, by fixing  $T$ and $s$ to 0.5 and varying $r$ and $t$, MCMS can be achieved as can be seen from Fig. \ref{fig:inq-sgad-t-r}.

\section{Conclusions}

We show that the effect of Unruh acceleration on a qubit is similar to
the action  of an amplitude  damping channel, where  the damping
parameter is  determined by  the acceleration, and  thus by  the Unruh
temperature.   Since   the  amplitude  damping  channel   arises  from
interaction with  a vacuum, this  suggests that the  Unruh temperature
should be regarded as a  factor that modifies the system's interaction
strength rather  than as the  temperature of the  damping environment.
 We study how environmentally induced
 decoherence modifies the effect of the Unruh channel,
essentially by investigating the degradation of quantum information,
as quantified by measures such as nonlocality, mixed-state entanglement,
teleportation fidelity, coherence and a discord-like quantity.
The differing effects of an  environment that
interacts dissipatively or non-dissipatively are noted.   
In particular, the latter is shown to
lead to a non-Pauli  channel of rank 3. Further,  useful  parameters characterizing channel performance such as gate and  channel fidelity 
are applied here to the Unruh channel, both with and without external influences. Squeezing is shown to be a useful resource in a number of scenarios.
 We hope this work is a contribution towards the development of
relativistic quantum technologies.

\end{document}